\RequirePackage{amsmath}
\documentclass[runningheads]{llncs}
\usepackage[T1]{fontenc}
\usepackage{graphicx}
\usepackage{listings}
\usepackage{xcolor}
\usepackage{flushend} 
\usepackage{tikz}
\usetikzlibrary{shapes.geometric, arrows, positioning, calc}

\definecolor{linkc}{rgb}{0.1,0.1,.8}
\definecolor{darkgreen}{rgb}{0,0.5,0}
\definecolor{midblue}{rgb}{0,0,0.7}
\usepackage[bookmarks=false,colorlinks=true,citecolor=linkc,linkcolor=linkc,filecolor=linkc,urlcolor=linkc]{hyperref}
\usepackage{url}

\usepackage{cleveref}
\usepackage{pgfplots}
\usepackage{makecell}
\usepackage{multirow}
\usepackage{comment}

\newcommand{\cuda}[0]{\textbf{\textsc{cuda}}}
\newcommand{\cgbn}[0]{\textbf{\textsc{cgbn}}}
\newcommand{\cudaQuad}[0]{\textbf{\textsc{Cu-Q}}}
\newcommand{\cudaFft}[0]{\textbf{\textsc{Cu-F}}}
\newcommand{\futharkQuad}[0]{\textbf{\textsc{Fu-Q}}}
\newcommand{\futharkFft}[0]{\textbf{\textsc{Fu-F}}}
\newcommand{\illegal}[0]{\textbf{\textcolor{red}{-{}-{}-{}-}}}

\newcommand{\kw}[1]{\mbox{\texttt{\bfseries{#1}}}}

\lstdefinelanguage{futhark}
{
  morekeywords={
    alloc,
    scan,
    map,
    map2,
    map3,
    map4,
    concat,
    def,
    do,
    else,
    f32,
    u32,
    bool,
    for,
    fun,
    if,
    in,
    include,
    let,
    loop,
    struct,
    then,
    type,
    val,
    while,
    with,
    mapnest,
    module,
    where,
    sort,
    scratch,
    multired,
    reverse,
    zip3,
    unzip3,
    copy,
    gather,
    withacc,
    sum
  },
  sensitive=true, 
  morecomment=[l]{--}, 
  morecomment=[s]{\{-}{-\}}, 
  morestring=[b]", 
  literate={\\}{\fn}{1} {->}{$\rightarrow$}{1} {<-}{$\leftarrow$}{1} {|>}{$\pipe$}{1},
}

\usepackage{xcolor}
\definecolor{eclipseBlue}{RGB}{42,0.0,255}
\definecolor{eclipseGreen}{RGB}{63,127,95}
\definecolor{eclipsePurple}{RGB}{127,0,85}

\newcommand{\pipe}{\ensuremath{\triangleright}}

\pgfplotsset{compat=1.18}

\newcommand{\mathword}[1]{\ensuremath{\text{\textit{#1}}}}
\newcommand{\codeword}[1]{\ensuremath{\text{\texttt{#1}}}}
\newcommand{\spacetune}[1]{#1}
\newcommand{\lstFigSize}{\scriptsize}
\newcommand{\lstDisplaySize}{\footnotesize}
\lstdefinestyle{lstFigStyle}{
   xleftmargin=2.5em,
   mathescape=true,
   escapechar=@,
   basicstyle=\lstFigSize\ttfamily,
   keywordstyle=\lstFigSize\color{midblue}\ttfamily\bfseries,
   commentstyle=\lstFigSize\slshape\color{darkgreen},
   numbers=left,
   numberstyle=\lstFigSize\rmfamily
}  
\lstdefinestyle{lstDisplayStyle}{
   xleftmargin=\parindent,
   mathescape=true,
   escapechar=@,
   basicstyle=\lstDisplaySize\ttfamily,
   keywordstyle=\lstDisplaySize\color{midblue}\ttfamily\bfseries,
   commentstyle=\lstDisplaySize\slshape\color{darkgreen},
   numbers=none
}  
\begin{document}
%

\title{GPU Implementations for Midsize Integer Addition and Multiplication}
%
%
\author{Cosmin E. Oancea\inst{1}\orcidID{0000-0001-5421-6876} \and\\
Stephen M. Watt\inst{2}\orcidID{0000-0001-8303-4983}}
%
%
%
\institute{DIKU, University of Copenhagen, Copenhagen 2100, Denmark
\email{cosmin.oancea@di.ku.dk}\\
\and
Cheriton School of Computer Science, University of Waterloo, Canada\\
\email{smwatt@uwaterloo.ca}}
\maketitle              
\begin{abstract}
This paper explores practical aspects of using a high-level functional language for GPU-based arithmetic on ``midsize'' integers.   
By this we mean integers of up to about a quarter million bits, which is sufficient for most practical purposes.  The goal is to understand whether it is possible to 
support efficient nested-parallel programs with a small, flexible code base.
We report on GPU implementations for addition and multiplication of integers that fit in one CUDA block, thus leveraging temporal reuse from scratchpad memories.
Our key contribution resides in the simplicity of the proposed solutions:
We recognize that addition is a straightforward application of scan,
which is known to allow efficient GPU implementation. 
For quadratic multiplication we employ a simple work-partitioning strategy that offers good temporal locality. For FFT multiplication,
we efficiently map the computation in the domain of integral fields
by finding ``good'' primes that enable almost-full utilization of machine words. 
In comparison, related work uses complex tiling strategies---which feel too big a hammer for the job---or uses the computational domain of reals, which may degrade the magnitude of the base in which the computation is carried. 
We evaluate the performance in comparison to the state-of-the-art CGBN library, authored by NvidiaLab, and report that our CUDA prototype outperforms CGBN for integer sizes higher than $32$K bits, while offering comparable performance for smaller sizes.  Moreover, we are, to our knowledge, the first to report that FFT multiplication outperforms the classical one on the larger sizes that still fit in a CUDA block.
Finally, we examine Futhark's strengths and weaknesses for efficiently supporting such computations and find out that the significant overheads and scalability issues of the Futhark implementation are mainly caused by the absence of a compiler pass aimed at efficient sequentialization of excess parallelism.
%
%

\end{abstract}
\newpage
\section{Introduction}

The work presented in this paper is ultimately aimed at extending already-parallel programs, written in high-level languages such as Futhark~\cite{futhark-pldi}, with support for multi-precision arithmetic that runs efficiently on highly-parallel hardware such as GPGPUs.  Accelerating such computations would benefit algorithms from various disciplines, such as computer algebra and cryptography.

The current work focuses on addition and multiplication of midsized integers, by which we mean integers of up to about a quarter million bits.  
These   suffice for most applications and, importantly, they 
fit in one \cuda~block of threads, or in a ``work group'' in OpenCL terminology.
This hardware level attention allows the implementation to leverage temporal reuse from fast (register+scratchpad) memory, which has significantly lower latency than global memory. In our experiments on Nvidia hardware, this range covers integers as large as about a quarter million bits.  

Normally, writing GPU software attending to this sort hardware consideration is time consuming, detailed and error-prone, with inflexible code using specific machine instructions for a particular platform. Our main interest has been to determine whether we may use a high level language to write GPU code that is elegant and flexible while being sufficiently efficient.  We have found this is readily achieved.   We have not at this stage been concerned with determining the best algorithms for different integer sizes (i.e. classical, versus Karatsuba, versus FFT), nor with using every trick to maximize performance (e.g. Montgomery representation for FFT).

\paragraph{Related Work}
The most closely related work is the ``Cooperative Groups Big Numbers'' (\cgbn) library~\cite{nvidialab:coopbignum}, authored by NvidiaLab, that offers a high-performance implementation for integers up to $32$K bits. The key technique used by \cgbn~to achieve top performance is to map an instance of integer computation on at most one warp of threads in order to leverage specialized Nvidia hardware that allows values to be communicated directly between registers (of the warp), i.e., without passing through scratchpad (shared) memory buffers that have significantly higher latency.  
Other related work aimed at integer in the target range implement
\begin{itemize}
\item the carry propagation of addition by mapping complex VLSI designs of hardware adders into equivalent GPU operations~\cite{JHPCA-FFT-10bits,ParProcLetters-AddSubMul-early2010},
\item classical (quadratic time) multiplication using tiling strategies~\cite{JHPCA-FFT-10bits}, but which require atomic updates to shared memory, or/and 
\item Strassen's algorithm~\cite{strassen1971schnelle} for log-linear-time multiplication by applying FFT in the domain of reals~\cite{cuFFT-base10,JHPCA-FFT-10bits}, which requires the computation to be carried in bases unfriendly to the underlying machine arithmetic (e.g., base $10$).
\end{itemize}
There have been other, earlier implementations of multiple precision integer arithmetic on CUDA, but with different emphasis:
\begin{itemize}
    \item CAMPARY~\cite{Campary1,Campary2} is a C library based on floating point arithmetic. 
    The main point is to  extend precision  by representing real numbers as the unevaluated sum of several standard machine precision floating-point values.
    \item CUMP~\cite{Cump1,Cump2} is an older work with a primary purpose of accessing CUDA with GMP.  Its stated objective was to outperform the GARPREC library~\cite{GARPREC}.
    Other similar works were ancestors of the \cgbn{} package, with which we compare directly.
    \item Isupov~\cite{Isupov-GRNS} uses interval techniques to augment residue number arithmetic for operations that rely on magnitude for numbers with upto 4096 bits.
    \item
Chen et al~\cite{Moreno-GPU-FFT} consider FFT using prime fields with generalized Fermat prime characteristics of size 504 and 992 bits to handle integers of practically unbounded size.   These numbers will generally span multiple \cuda{} blocks, hence the work is mostly concerned with making sure that access patterns enable coalesced access to global memory.  They aim to improve spatial locality, but not temporal locality (re-use from shared memory).
\end{itemize}

We have recently assisted to a proliferation of Python-embedded DSLs aimed at supporting ML practitionars---such as Tensorflow~\cite{abadi2016tensorflow}, PyTtoch~\cite{paszke2019pytorch} and Jax~\cite{frostig2018compiling}. Similarly, we envision that a practically important direction refers to interfacing languages supporting multi-precision arithmetic with mainstream computer-algebra systems such as Maple~\cite{Maple} and Mathematica~\cite{Mathematica}, thus enabling both ease of scripting and high performance. Such solutions can build on prior interoperability work aimed, for example, at supporting automatic differentiation~\cite{PyTorch-Julia-AD,mixed-lang-AD} or parameteric polymorphism~\cite{mapal_synasc,gidl_oopsla,alma:ISSAC} across language boundaries.


\enlargethispage{\baselineskip}
\paragraph{Contributions}
This paper presents \cuda~and Futhark implementations for integer addition and multiplication, including both quadratic and Strassen's log-linear time algorithms.
The main contribution of our solutions mainly resides in choosing the simplest tool that does the job: We recognize that addition is a straightforward application of scan~\cite{segScan}---a basic-block of parallel programming---for which efficient GPU implementations are folklore~\cite{single-pass-scan}. 
For classical multiplication we use a simple partitioning technique that assigns to each thread a load-balanced computation of entire elements of the result, so that updates do not need to be atomic and are performed directly in low-latency registers.
For Strassen (FFT) multiplication, we conduct the computation in the integral domain by using computer-algebra reasoning to find good prime fields that maximize the utilization of machine-supported arithmetic. This allows for example to represent integers using $15$ bits of each half word or $31$ bits of each word.

In comparison to \cgbn, our implementation does not relies on specialized hardware instructions and, we surmise, is likely to translate the performance to other hardware from different vendors, such as AMD. Instead of mapping integer operations to execute at warp level, our implementation allows them to occupy as much as an entire \cuda~block of threads, and relies on the classical technique of efficiently sequentializing parallelism in excess\footnote{In simple words, this refers to having one thread compute in a sequential-efficient fashion several elements of the result instead of just one.} to amortize each access to shared memory across several register accesses.

We evaluate the performance of our implementations in comparison with \cgbn---on programs performing one addition, one multiplication and fusion of such operations---and report that \cgbn~is faster on integer ranges up to $2^{13}$ bits, but our \cuda~implementation gains the upper hand on ranges of $2^{15}-2^{16}$ bits, and outperforms \cgbn~on integers consisting of $2^{17}$ and $2^{18}$ bits.  In fairness, \cgbn~offers near-perfect scalability on fused operations.   

Our Futhark implementation exhibits significant overheads in comparison to our \cuda~prototype. The performance bottlenecks are caused by the absence of a compiler pass that automatically performs efficient sequentialization. Implementing it by hand is possible in Futhark, but still sub-optimal in several ways: {\em First}, intermediate results are always mapped by the compiler to shared-memory buffers and there does not exist a way for the programmer to change the mapping to register memory. {\em Second}, the shared-memory mapping may restrict the maximal size of the integer that fits in a \cuda~block. {\em Finally}, the user code implementing efficient sequentialization is likely to degrade the performance of other semantically-equivalent code versions that, for example, support execution even when the integer is too big to fit in one \cuda~block. \smallskip 

In summary, the contributions of this paper are:\vspace{-1ex}
\begin{itemize}
\item a demonstration that high level languages (\texttt{C++} and Futhark) can be used to implement big integer arithmetic concisely and efficiently for GPU computation,
\item simple and efficient GPU implementations for multi-precision addition and multiplication,
\item an experimental evaluation that demonstrates significant performance gains in comparison to \cgbn~library on integer sizes in the range of $2^{15}$ to $2^{18}$ bits,
\item to our knowledge, the first demonstration that FFT-based multiplication outperforms an efficient implementation of the quadratic algorithm on sizes that fit in a \cuda~block (by factors as high as $5\times$ on the largest size),
\item a presentation that (we hope) allows to reproduce the implementations directly from the information in the paper. 
\end{itemize}


\paragraph{Outline}
This paper is structured in a straightforward fashion: Sections~\ref{sec:badd},~\ref{sec:bquadmult}~and~\ref{sec:bfftmult} present our implementations of addition, quadratic and log-linear time multiplication, respectively. Section~\ref{sec:futhark} discusses the strengths and weaknesses of the current Futhark compiler infrastructure for supporting multi-precision arithmetic.  Section~\ref{sec:experiments} reports experimental results, and
Section~\ref{sec:conclusions} concludes.



\section{Integer Addition}
\label{sec:badd}

We represent a big unsigned integer---referred from now on as an 
integer---as an array $a$ containing $M$ elements
of type $uint$, which allows to store $M\cdot 8 \cdot \codeword{sizeof(uint)}$
bits. The elements can be seen as the coefficients of a 
polynomial in (base) $x = 2^{8\cdot \codeword{sizeof(uint)}}$, i.e.,
$a = a_0 + a_1\cdot x + \dots a_{M-1}\cdot x^{M-1}$.
In our implementation of addition and multiplication, the
result has the same length and element type as the input integers, 
e.g., $add ~ : ~ [M]uint ~ \rightarrow ~ [M]uint ~ \rightarrow ~ [M]uint$.

\begin{figure}[t]
\hspace{0.5ex}%
\begin{center}
\begin{minipage}{0.375\textwidth}
    \includegraphics[width=\textwidth]{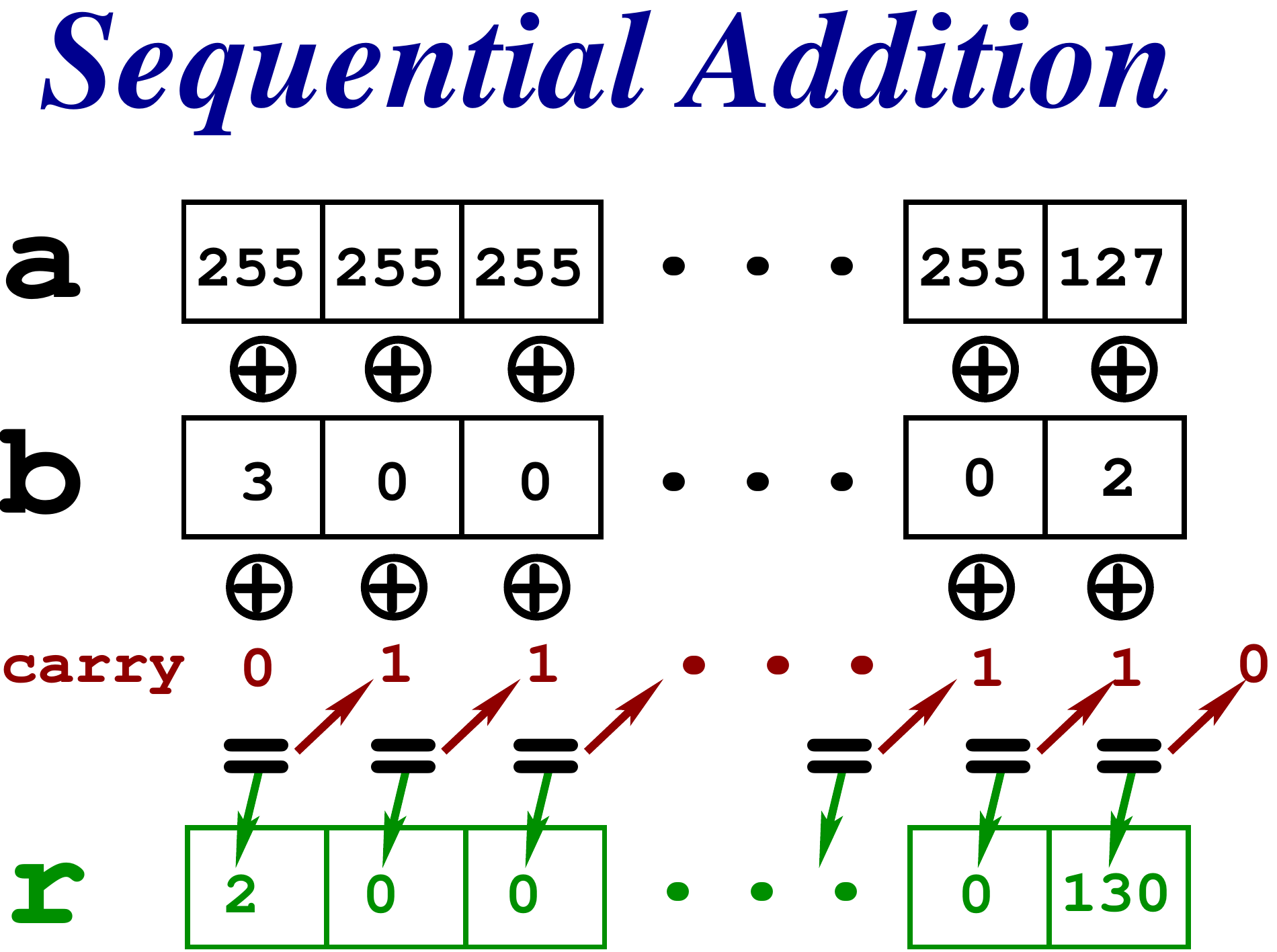} 
    \vspace{0.01ex}
\end{minipage}\hspace{3ex}%
\begin{minipage}{0.485\textwidth}
\includegraphics[width=\textwidth]{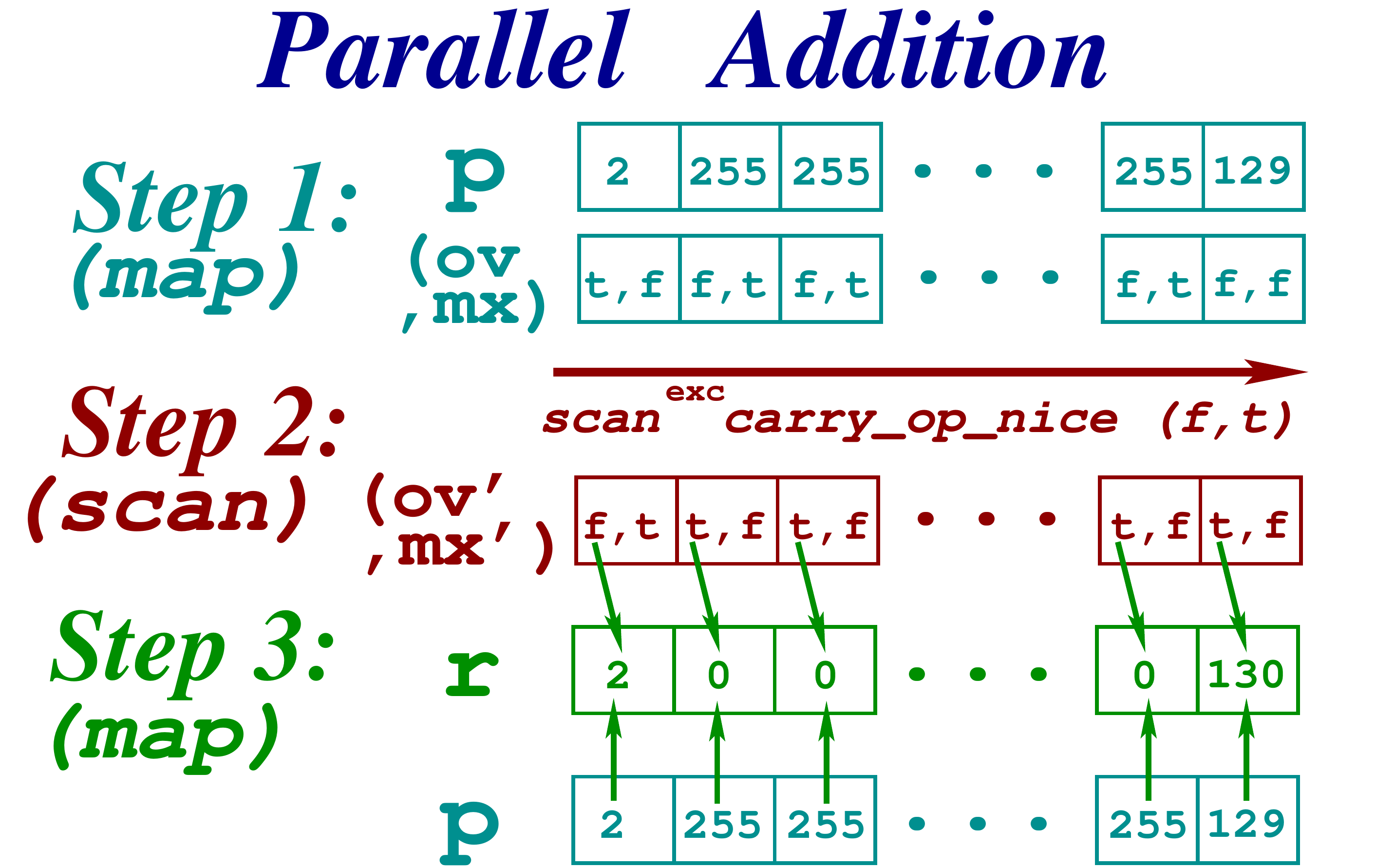}
\end{minipage}
\end{center}
    
\caption{Sequential and parallel procedures for addition {\tt{}a+b} base $2^8$: linear vs log time.}
\label{fig:add-seq-intuition}
\end{figure}

Adding two such integers can be accomplished by a (well known) 
iterative procedure, illustrated on the left side of 
figure~\ref{fig:add-seq-intuition},
that adds (in a bigger type of double size) the corresponding elements
of $a$ and $b$ together with the carry from the previous operation, and
then it computes the result element and the carry for the next iteration
as the remainder and quotient of the division of the sum to the integer's 
base. 
    This is commonly implemented by performing the additions within
    the domain of $uint$ and by checking for overflow.

The procedure described above is inherently sequential, since the carry
computation gives rise to a cycle of cross-iteration true dependencies.
Furthermore, figure~\ref{fig:add-seq-intuition} shows a pathological
case in which the carry of the first addition is propagated all the
way to the last element of the result, which, at first sight, would
seem that it does not allow any chunks of computations to be performed
in parallel, i.e., independently of each other.

However, as reported in~\cite{carry-by-scan} and illustrated on the right
side of figure~\ref{fig:add-seq-intuition}, a data-parallel implementation
can be straightforwardly obtained by reasoning in terms of the basic
blocks of parallel programming---particularly scan~\cite{segScan,BlellochCACM96NESL}, a.k.a., 
prefix sum. The procedure requires three parallel steps:
\begin{itemize}
    \item[(1)] a \lstinline{map} operation that independently sums
            up corresponding elements within the $uint$ domain,
            and computes a partial result $uint \ p_i \ = \ a_i + b_i$,
            together with two booleans $ov_i$ and $mx_i$ that record
            whether (i) the addition has overflow, i.e., there is a carry,
            and (ii) the result of the addition is the highest
            element of $uint$, i.e., $ov_i = (p_i < a_i)$ and 
            $mx_i = (p_i == uint.highest)$.
    \item[(2)] an exclusive \lstinline{scan} that combines the 
                overflow-highest pairs across elements,~and
    \item[(3)] another \lstinline{map} that adds the partial result
                obtained in step (1) with the carry (overflow) result
                obtained from the scan in step (2).
\end{itemize}

\noindent Before zooming on step (2), we recall the type and semantics of {\em exclusive} scan:
\[ 
    \mathrm{scan}^{\mathword{exc}} \ : \ (\tau \rightarrow \tau \rightarrow \tau) \ \rightarrow \ \tau \ \rightarrow [n]\tau \ \rightarrow \ [n]\tau \\
\]
\[
    \mathrm{scan}^{\mathword{exc}} \ \odot \ e_\odot \ [a_1,\ldots,a_n] \ \equiv \ [e_\odot, \ a_1, \ a_1 \odot a_2, \ \ldots, \ a_0 \odot \ldots \odot a_{n-1}]
\]
where $\odot$ is an arbitrary {\em associative} binary operator with
neutral element $e_\odot$. Scan has parallel work $O(n)$ and depth $O(\log n)$.
Carries can be propagated by means of a scan with an operator that
is expressed in a friendly way as below: 
\spacetune{\enlargethispage{2\baselineskip}}
\vbox{
\begin{lstlisting}[language=futhark,style=lstDisplayStyle]
-- neutral element is (false, true)
def carry_op_nice (ov1: bool, mx1: bool)
                  (ov2: bool, mx2: bool) : (bool, bool) =
    ( (ov1 && mx2) || ov2,  mx1 && mx2 )
\end{lstlisting}
}
\noindent Regarding {\tt(ov1,mx1)} as the current accumulator
and {\tt(ov2,mx2)} as the current element of the scanned array,
the (sequential) rationale of the operator is:
\begin{itemize}
    \item An overflow needs to be carried to the next position  
        if either (i) the current addition has produced an overflow,
        i.e., {\tt ov2} holds, or (ii) the current addition has 
        resulted in a maximal element of {\tt uint} {\em and}
        the current accumulator signals an overflow, i.e., \verb+(ov1 && mx2)+ holds.
    \item All elements scanned so far are the maximal element
        of $uint$ {\em iff} the current accumulator and the current
        element are maximal,\footnote{
            While {\tt mx1} is not actually used in the sequential
            interpretation, it is essential for the parallel execution,
            which uses a circuit network of depth $O(\log n)$ to compose
            the results of smaller scanned segments.}
            i.e., {\tt mx1 \&\& mx2}.
\end{itemize}
Topalovic, Restelli-Nielsen and Olesen prove that the operator is associative~\cite{carry-by-scan},
and that it accepts \lstinline{(false,true)} as its neutral element,
which is easy to see.

\begin{figure}[t]
\begin{lstlisting}[language=futhark,style=lstFigStyle]
-- neutral element is 2 for both carry_op_eff and carry_op_sgm
def carry_op_eff (c1: u32) (c2: u32) =
    (c1 & c2 & 2) | ( ((c1 & (c2 >> 1)) | c2) & 1 )

def carry_op_sgm (c1: u32) (c2: u32) =
    if (c2 & 4) != 0 then c2
    else ( (carry_op_eff c1 c2) | ( (c1 | c2) & 4 ) )

-- computes IPB instances but without efficient sequentialization
def badd [IPB][M] (as : [IPB*M]u32) (bs : [IPB*M]u32) : [IPB*M]u32 =
    let f a b i =
        let p = a + b 
        let b = ((u32.bool (i % M == 0)) << 2)
              | ((u32.bool (p == u32.highest)) << 1)
              | (u32.bool (p < a))
        in  (p, b)
    let (part_res, carry_elms) = map3 f as bs (0..<IPB*M) |> unzip @\label{line:badd-step1}@
    
    let carries = scan$^{exc}$ carry_op_sgm 2 carry_elms -- carry propagation @\label{line:badd-step2}@
    
    let g r c = r + u32.bool ( c & 1 == 1 )
    in  map2 g part_res carries @\label{line:badd-step3}@

def bbadd [N][IPB][M] 
        (ass : [n][IPB*M]u32) (bss : [n][IPB*M]u32) : [n][IPB*M]u32 =
    map2 badd ass bss @\label{line:badd-batch}@
\end{lstlisting}\vspace{-2ex}
\caption{Futhark pseudocode for performing a batch of {\tt IPB} 
        additions of big numbers, each represented as an array of
        {\tt M} $32$-bit unsigned integers. Function {\tt badd} is supposed to be mapped at CUDA-block level (where arrays are mapped to scratchpad memory). 
        Efficient sequentialization is not shown, albeit it is critical for good performance.
        }
\label{fig:badd}
\end{figure}

Figure~\ref{fig:badd} shows a more involved Futhark implementation,
denoted {\tt badd}, which is intended to be mapped at the CUDA-block
level of parallelism, such that all intermediate arrays are maintained
and accessed from fast (scratchpad or register) memory. 
The pseudocode specializes for simplicity {\tt uint} to $32$-bit unsigned
integer (\lstinline{u32}) and
computes {\tt IPB} instances of additions in a CUDA block---e.g., 
to optimize the case when {\tt M} is too small for a good block size.
Details are:
\begin{itemize}
\item {\tt carry\_op\_eff} is very similar to {\tt carry\_op\_nice},
    excepts that it packs the tuple of boolean values in the last two
    bits of an \lstinline{u32} value. The rationale is that this 
    (i) requires only one scratchpad buffer (instead of two),
    and (ii) GPU hardware is optimized for \lstinline{u32} accesses; 
    otherwise one can also use \lstinline{u8}.
\item However, since we aim to compute {\tt IPB} (independent) instances
    within a CUDA block, step (2) needs to perform a segmented scan instead
    of a scan. This is typically achieved by lifting the scan's operator
    to operate over tuples formed by the original datatype and 
    a boolean which, when set, indicates the start of a segment.
    {\tt carry\_op\_sgm} encodes the lifted operator of the segmented
    scan by encoding the start of the segment in the third-last bit.
\item the implementation of {\tt badd} follows the three parallel
    steps mentioned before: the first corresponds to \lstinline{map3 f}
    at line~\ref{line:badd-step1}, which computes the partial result and the input to the
    (segmented) scan, the second step corresponds to 
    \lstinline{scan carry_op_sgm 2} at line~\ref{line:badd-step2}, which propagates the
    carries, and the third step---\lstinline{map2 g} at line~\ref{line:badd-step3}---adds 
    the resulted carries to each partial result.
\item the {\tt bbadd} function performs an arbitrary batch ($N$) of
    {\tt badd} computations, hence the \lstinline{map2} at line~\ref{line:badd-batch}
    is intended to be mapped on CUDA's grid of blocks.
    
\end{itemize}

A final optimization that we apply (not shown) is the classical 
{\em efficient sequentialization of excess parallelism}. 
In our case, this corresponds to having each thread process independently
a parametric (statically-known, smallish) number of elements, rather
than just one, as a way of reducing the inter-thread communication
overhead for operators such as scan and reduce.\footnote{
Our CUDA implementation of block-level scan and reduce follows the
standard strategy~\cite{single-pass-scan} that (de)composes 
the implementation hierarchically, at each level of the hardware: 
CUDA thread, warp and block level.}

The practical manner of implementing efficient sequentialization in
CUDA is by mapping logical arrays to register (thread-private) memory 
whenever possible, and by using shared memory preferably only as staging 
buffers---e.g., for copying in coalesced way to/from global
to register memory or for storing intermediate results (of reduced size)
produced in the internal implementation of \lstinline{reduce} and 
\lstinline{scan}.
(Of course, some operations force manifestation of logical arrays in shared
memory, e.g., when the same element is read by multiple threads, or in the
presence of gather/scatter operations that access statically unknown indices.)
This strategy has well known advantages:
\begin{itemize}
\item it allows to maximize the utilization of both register and shared memory,
\item it promotes accesses from registers, which has lower latency, do not
suffer bank conflicts, and are more numerous (in terms of bytes per thread)
than shared memory,
\item it minimizes the live range of shared-memory buffers, thus promoting
their reuse, while register usage is automatically optimized by register
allocation.
\end{itemize}
In summary, efficient sequentialization is a performance critical optimization
that reduces inter-thread communication and the latency of memory accesses,
generating significant speedups, higher than $2\times$ in cases. More importantly,
it enables the implementation to {\em efficiently support larger integers}, 
since in our context, their size is tied with that of the CUDA block: 
On the one hand, the quantity of resources utilized by a CUDA block is typically 
proportional with its size, hence a suboptimal mapping will lower the size
of the CUDA block that can be launched and thus the magnitude of the integer.  
One the other hand, CUDA blocks is hard constrained to a maximal number of 
$1024$ threads, hence a one-to-one mapping would limit the integers to 
$[1024]uint$, while an efficient sequentialization factor $Q = 8$ would support 
$[8196]uint$.
In principle, for addition, the integer size is constrained by registers, not 
by shared memory, because all logical arrays can be mapped to register memory.

\section{Classical, Quadratic Multiplication}
\label{sec:bquadmult}

Discussion is organized as follows: Section~\ref{subsec:insights}
provides the high-level rationale of our implementation strategy in comparison with the popular approach of applying tiling to optimize convolution-based code. Section~\ref{subsec:birds-eye-view} gives the birds-eye-view of our CUDA implementation, and section~\ref{subsec:conv-implem} zooms in on the main computational step (the convolution). 

\subsection{Key Insights}
\label{subsec:insights}

The classical algorithm for multiplication corresponds to the formula:
\begin{equation}
\label{eqn:clasical-mul}
C_{k} = \sum_{\substack{i+j = k \\ 0\leq i,j,k < M}} A_i \cdot B_j
\end{equation}
\noindent which assumes that the element type $uint'$ of the result $C$ 
is large enough to prevent overflow. Denoting with $uint$ the element type
of $A$ and $B$, practical implementations commonly perform the product 
$A_i \cdot B_j$ inside a type $ubig$ which has double the size of $uint$, 
thus guaranteeing no overflow, and represent the element type of $C$
as a tuple $(ubig, uint32)$ in which the second term denotes the carry
that accounts for the potential overflow of summation. Alternatively, one
may use a triple $(uint,uint,uint32)$ in which the first two terms 
correspond to the low and high part of $ubig$.   

This section is aimed to highlight the key differences between implementation
choices and thus will work directly with formula~\eqref{eqn:clasical-mul} and
ignore the overflow details.

\subsubsection{Tiling Approach.}
%
%
Related approaches~\cite{cuFFT-base10,JHPCA-FFT-10bits,ParProcLetters-AddSubMul-early2010} predominantly use block tiling to implement 
formula~\eqref{eqn:clasical-mul}; this results in C-like code similar to the
one below, which uses for simplicity the same tile size {\tt T} that is assumed
to evenly divide~{\tt M}:

\begin{lstlisting}[language=C++,style=lstDisplayStyle]
for(int k = 0; k < M; k++) C[k] = 0;

for (int ii = 0; ii < M); ii+=T)  // mapped on CUDA Grid.y
  for (int jj = 0; jj < M; jj+=T) // mapped on CUDA Grid.x
    for (int i = 0; i < T; i++)   // mapped on CUDA Block.y
      for (int j = 0; j < T; j++) // mapped on CUDA Block.x
        if (ii+jj + i+j < M) 
          C[ii+jj + i+j] += A[ii+i] * B[jj+j];
\end{lstlisting}

The tiled code minimizes the temporal reuse distance of the
accesses to $A$, $B$ and $C$. For example, the read indices of $A$
are invariant to loop {\tt j} (and {\tt jj}). It follows that
the slice of {\tt A[ii : ii+T]} can be remapped to a scratchpad
memory buffer of length {\tt T} just inside the {\tt jj} loop 
and reused from there within the body of the loop, i.e., one
access to global memory is amortized across $T$ accesses to
scratchpad memory. Similar thoughts apply to arrays $B$ and $C$.
However, the loop nest above is a generalized reduction~\cite{genred-crummey,CosPLDI},
whose parallelization requires inter-thread communication, because
the same element of {\tt C} may be updated by different threads,
hence the additive updates need to be atomic.

\begin{figure}[t]
\begin{lstlisting}[language=c++,style=lstFigStyle]
template<uint, ubig, uint32_t M, uint32_t T> __global__ 
void bmulTiled ( uint* Aglb, uint* Bglb, ubig* Cglb ) {
    __shared__ uint Ash[T], uint Bsh[T]; __shared__ ubig Csh[2*T];
    int ii = blockIdx.y*T, i = threadIdx.y; // 0 <= i < T
    int jj = blockIdx.x*T, j = threadIdx.x; // 0 <= j < T
    // copy A and B from global to shared memory & initialize Csh
    if(threadIdx.y == 0) { 
        Ash[j] = Aglb[ii+j]; Csh[j    ] = 0;
        Bsh[j] = Bglb[jj+j]; Csh[j + T] = 0;
    }
    __syncthreads();
    if(ii+jj + i+j < M) {
        ubig prod = ((ubig)Ash[i]) * ((ubig)Bsh[j]);
        atomicAdd(&Csh[i+j], prod);  // atomic in shared memory @\label{line:mtk-atomic1}@
    }
    __syncthreads();
    int tid = i*T + j;
    if(tid < 2*T && ii+jj + tid < M) // atomic in globla memory
        atomicAdd( &Cglb[ii+jj + tid], Csh[tid] ); @\label{line:mtk-atomic2}@
}
\end{lstlisting}
\caption{Sketch of a simple CUDA kernel for the tiled version of quadratic multiplication.}
\label{fig:classic-mul-tiled-ker}
\end{figure}

Figure~\ref{fig:classic-mul-tiled-ker} sketches a toy CUDA kernel implementing the tiled version, which assumes a two-dimensional block of size {\tt{}T$\times$T}. This approach optimizes temporal locality and enables maximal parallelism but has two shortcomings:
\begin{itemize}
\item {\tt C} is not only maintained in shared memory, which has higher latency then registers, but its updates use {\em expensive} atomic operations: {\tt{}T$\cdot$T} times
    from shared memory (line~\ref{line:mtk-atomic1} in figure~\ref{fig:classic-mul-tiled-ker}) and {\tt$2\cdot$T} times from global memory~(line~\ref{line:mtk-atomic2}).
\item 
    it prevents producer-consumer fusion---e.g., with following additions and
    multiplications---because the atomic add in global memory requires a global
    barrier across all blocks, which is not possible in CUDA other than by
    ending the execution of the current kernel.
\end{itemize}

\subsubsection{Load Balanced Partitioning of the Result Across Threads.} 

\begin{figure}[t]
\[
\begin{array}{lll@{~~~~~}r}
\textcolor{red}{C_0}        & \ = \ & A_0 \cdot B_0 & \textcolor{red}{\mbox{\tt 1 term~}}\\
\textcolor{blue}{C_1}      & \ = \ & A_0 \cdot B_1 + A_1 \cdot B_0 & \textcolor{blue}{\mbox{\tt 2 terms}}\\
\textcolor{darkgreen}{C_2}       & \ = \ & A_0 \cdot B_2 + A_1 \cdot B_1 + A_2 \cdot B_1 & \textcolor{darkgreen}{\mbox{\tt 3 terms}}\\
\ldots                      &       & \ldots \\
\textcolor{darkgreen}{C_{M-3}}   & \ = \ & A_0 \cdot B_{M-3} + \ldots + A_{M-3} \cdot B_{0} & \textcolor{darkgreen}{\mbox{\tt{}M-3 terms}}\\
\textcolor{blue}{C_{M-2}}  & \ = \ & A_0 \cdot B_{M-2} + \ldots + A_{M-3} \cdot B_{1} + A_{M-2} \cdot B_0 & \textcolor{blue}{\mbox{\tt{}M-2 terms}}\\
\textcolor{red}{C_{M-1}}    & \ = \ & A_0 \cdot B_{M-1} + A_1 \cdot B_{M-2} + \ldots + A_{M-2} * B_{1} + A_{M-1}\cdot B_{0} & \textcolor{red}{\mbox{\tt{}M-1 terms}}\\
\end{array}
\]
\caption{A load-balanced embarrassingly parallel partitioning is to assign thread $0$ to compute $C_0$ and $C_{M-1}$, thread $1$ to compute $C_1$ and $C_{M-2}$, thread $2$ to compute $C_2$ and $C_{M-3}$, and so on. All threads perform a total $M$ multiply-fused add operations.}
\label{fig:partitioning-res-across-threads}
\end{figure}

We choose instead a strategy that partitions the elements of the result in a manner that is
load balanced, and assigns the computation of an entire partition to the same thread.
%
%
Figure~\ref{fig:partitioning-res-across-threads} illustrates the partitioning that computes two elements of the result with each thread, i.e., the identically coloured elements are computed by the same thread and they require the same number ($M$) of terms. One could increase the sequentialization degree, denoted $Q$, by computing $Q=4$ or $Q=8$ elements of the result per thread, which would result in $2\cdot M$ and $4\cdot M$ terms computed by each thread, respectively. This strategy has the advantages that it:
\begin{itemize} 
\item allows the result $C$ to be mapped to register memory, and to be computed in an embarrassingly parallel fashion, while enabling efficient sequentialization,
\item allow multiple addition/multiplication operations to be fused within a CUDA block, such that intermediate results are reused from fast memory.
\end{itemize}
The downside is that it sequentializes completely an entire parallel dimension of size {\tt M}. We found however that this is a small price to pay, given the advantages, especially when considering that 
\begin{itemize}
\item we aim at integrating such arithmetic inside programs that are already parallel, which makes it unlikely that the parallel hardware will be starved,
\item classical multiplication has suboptimal $O(M^2)$ work, in comparison with the $O(M ~\log M)$ FFT algorithm, hence it makes sense to use it as a niche specialization aimed at squeezing maximal performance from the hardware.
\end{itemize}

\subsection{Birds-Eye View of Implementation}
\label{subsec:birds-eye-view}

Figure~\ref{fig:classic-mul-regs} shows the CUDA implementation of the
entry function {\tt bmulRegs} that implements the quadratic multiplication
algorithm.  The implementation denotes by {\tt Q} half the sequentialization
factor, i.e., each thread computes {\tt $2\cdot$Q} elements of the result,
and assumes that {\tt $2\cdot$Q} evenly divides {\tt M}, which has been
(previously) ensured by padding the numbers to the closest multiple of
{\tt Q}. 

By this point the input has already been read in coalesced way
from global to register memory (not shown), hence the function's input
{\tt Areg} and {\tt Breg} and result {\tt Rreg} are stored in register 
memory.  {\tt Ash} and {\tt Bsh} are shared-memory staging buffers of 
length {\tt [IPB*M]$uint$}.   
As before, {\tt IPB} denotes the number of instances solved within a 
CUDA block, and the integer consists of {\tt M} elements of type
{\tt uint}.  Thus the size of the CUDA block size is: 
{\tt IPB*M / (2*Q)}.

\begin{figure}[t]
\begin{lstlisting}[language=C++,style=lstFigStyle]
template<class Base,uint32_t IPB,uint32_t M,uint32_t Q> __device__ void
bmulRegs( typename Base::uint*Ash,       typename Base::uint*Bsh,
          typename Base::uint Areg[2*Q], typename Base::uint Breg[2*Q],
          typename Base::uint Rreg[2*Q] 
) {
    using uint = typename Base::uint;
    using ubig = typename Base::ubig;
        
    // 1. copy from register to shared memory
    cpReg2Shm<uint, 2*Q>( Areg, Ash ); @\label{line:mul-reg-man1}@
    cpReg2Shm<uint, 2*Q>( Breg, Bsh ); @\label{line:mul-reg-man2}@
    __syncthreads();

    // 2. perform the convolution
    uint lhcs[2][Q+2];
    wrapperConv<uint, ubig, M, Q>( Ash, Bsh, lhcs ); @\label{line:mul-reg-conv}@
    __syncthreads();
        
    // 3. publish low parts & high and carry (hcs[:][Q:]) in Lsh & Hsh
    typename Base::uint *Lsh = Ash, *Hsh = Bsh;
    publishReg2Shmem<uint, M, Q>( lhcs, Lsh, Hsh );
    __syncthreads();
    
    // 4. load back to register and perform the addition of the carries.
    uint Lrg[2*Q], Hrg[2*Q];
    cpShm2Reg<uint, 2*Q>( Lsh, Lrg ); @\label{line:mul-regs-remap1}@
    cpShm2Reg<uint, 2*Q>( Hsh, Hrg ); @\label{line:mul-regs-remap2}@
    __syncthreads();    
    baddRegs<uint, M, 2*Q, Base::HIGHEST>( Lsh, Lrg, Hrg, Rreg ); @\label{line:mul-regs-baddRegs}@
}
\end{lstlisting}
\caption{Main CUDA wrapper function that computes quadratic integer multiplication.
    }
\label{fig:classic-mul-regs}
\end{figure}

The implementation manifests numbers $A$ and $B$ in shared memory (at lines~\ref{line:mul-reg-man1}-\ref{line:mul-reg-man2}), because computing the convolution (line~\ref{line:mul-reg-conv}) corresponding to formula~\eqref{eqn:clasical-mul} requires multiple threads to access same elements of $A$ and $B$.

The per-thread result of the convolution (line~\ref{line:mul-reg-conv}) is the array named {\tt lhcs} which has type $[2][Q+2]uint$. The rationale is that each thread processes two contiguous sub-partitions of $Q$ elements: one from the first half of the integer and its symmetric opposite across the midpoint. The result of each sub-partition is represented as an array 
of size $Q+2$ of $uint$s that addresses overflow concerns: 
\begin{itemize}
\item the first $Q$ elements are the low parts of the sequentially aggregated result,
\item the next element correspond to the high part of the aggregated result, and the last one to an additional carry (in case the high part overflows).
\end{itemize}

\begin{figure}[t]
\begin{center}
    \includegraphics[width=0.8\textwidth]{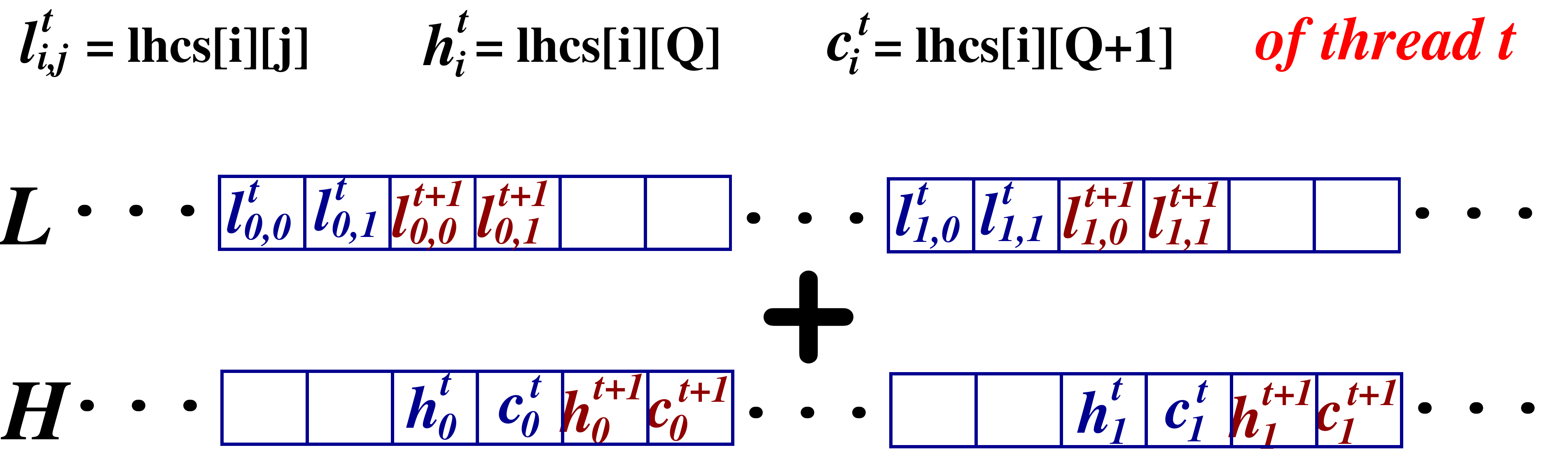} 
\end{center}
\caption{Illustrating the placement of thread-private array {\tt lhcs$: [2][Q+2]uint$} in shared-memory arrays $L$ and $H$ for $Q=2$. $L$ and $H$ are then added to complete the algorithm.}
\label{fig:place-lhcs}
\end{figure}

Next, the function {\tt publishReg2Shmem} manifests the per-thread
aggregated {\tt lhcs} result into two shared-memory buffers denoted
{\tt Lsh} and {\tt Hsh} in the manner depicted in figure~\ref{fig:place-lhcs}.
(Note that these are in fact reusing the shared-memory buffers of {\tt Ash}
and {\tt Bsh}, which are dead after convolution). Finally, the low, high
and carry parts are aggregated across threads by simply adding together, 
with the procedure from section~\ref{sec:badd}, the integers represented 
by the arrays $L$ and $H$: 
This is accomplished by the call to {\tt baddReg} at line~\ref{line:mul-regs-baddRegs}, which
requires its arguments to be remapped as registers (at lines~\ref{line:mul-regs-remap1} and~\ref{line:mul-regs-remap2}).

\subsection{Implementing The Convolution}
\label{subsec:conv-implem}

\begin{figure}
\begin{lstlisting}[language=C++,style=lstFigStyle]
template<class uint,class ubig, uint32_t M,uint32_t Q> __device__
void combine( ubig accums[Q], uint32_t carrys[Q], S lhcs[Q+2] ) {
    const uint32_t SHFT = 8 * sizeof(uint);
    lhcs[0] = (uint) accums[0];
    uint h_res = (uint) (accums[0] >> SHFT);
    uint c_res = carrys[0];
    
    for(int q=1; q<Q; q++) {
        uint l = (uint) accums[q];
        uint h = (uint) (accums[q] >> SHFT);
        lhcs[q] = l + h_res;
        h_res = h + (c_res + (lhcs[q] < l));
        c_res = carrys[q] + (h_res < h);
    }
    lhcs[Q]   = h_res;
    lhcs[Q+1] = c_res;
}

template<class uint, class ubig> __device__  void 
convIter( uint32_t i, uint32_t j, uint* Ash, uint* Bsh, 
          ubig& accum, uint32_t& carry
) {
    const uint32_t SHFT = 8*sizeof(S);
    uint accum_prev = (uint) (accum >> SHFT);
    accum += ((ubig)Ash[i]) * ((ubig)Bsh[j]); @\label{line:mul-conv-accum}@
    carry += (  ((uint)(accum >> SHFT)) < accum_prev ); @\label{line:mul-conv-carry}@
}

template<class uint,class ubig, uint32_t M,uint32_t Q> __device__
void convolution(uint32_t k1, uint* Ash, uint* Bsh, uint lhcs[Q+2]) {
    ubig accums[Q];  uint32_t carries[Q];
    for(int q=0; q<Q; q++) { 
        accums[q] = 0; carries[q] = 0; 
    }
    for(int kk = 0; kk <= k1; kk++) { @\label{line:mul-conv-qstart}@
        uint32_t i = kk;
        uint32_t j = k1 - i;
        for(int q=0; q<Q; q++)
          convIter<uint,ubig>(i,j+q, Ash,Bsh, accums[q],carries[q]); @\label{line:mul-conv-iter1}@
    }
    for(int q=1; q<Q; q++) {
        for(int i=0; i<Q-q; i++)
          convIter<uint,ubig>(k1+q,i,Ash, sh,accums[i+q], carries[i+q]); @\label{line:mul-conv-iter2}@
    } @\label{line:mul-conv-qend}@
    combine<uint,ubig,M,Q>(accums, carries, lhcs); @\label{line:mul-conv-combine}@
}       
template<class uint, class ubig, uint32_t M, uint32_t Q> __device__
void wrapperConv( uint* Ash0, uint* Bsh0, uint lhcs[2][Q+2] ) {
    const uint32_t offset = ( threadIdx.x / (M/(2*Q)) ) * M;
    uint *Ash = Ash0 + offset, *Bsh = Bsh0 + offset;
    uint32_t ltid = threadIdx.x % (M/(2*Q));
    
    convolution<uint,ubig,M,Q>( Q * ltid,     Ash, Bsh, lhcs[0]); //first half @\label{line:mul-conv-conv1}@
    convolution<uint,ubig,M,Q>( M-Q*(ltid+1), Ash, Bsh, lhcs[1]); //second half @\label{line:mul-conv-conv2}@
}        
\end{lstlisting}
\caption{CUDA code for computing the per-thread convolution.}
    \label{fig:classic-mul-conv}
\end{figure}

Figure~\ref{fig:classic-mul-conv} details the implementation of the
convolution step:
\begin{itemize}
\item Function {\tt wrapperConv} calls twice function {\tt convolution} at lines~\ref{line:mul-conv-conv1} and \ref{line:mul-conv-conv2}, each call processing one of the contiguous sub-partitions corresponding to $Q$ elements of the result,
\item Function convolution computes each of the $Q$ results independently in the two loops at lines~\ref{line:mul-conv-qstart}-\ref{line:mul-conv-qend}. The two loops are necessary since each element of the result requires incrementally more terms. 
\item Function {\tt convIter}, called at lines~\ref{line:mul-conv-iter1} and \ref{line:mul-conv-iter2} adds one more term to the result. We represent the result at this stage as a tuple between {\tt accum} of type $ubig$ (which has double the size of $uint$) and a {\tt carry} of type $uint32_t$. The new term $A_i \cdot B_j$ is computed in the $ubig$ type (line~\ref{line:mul-conv-accum}), which prevents overflow.  However the additive update of {\tt accum}  may result in overflow, which is accounted for at line~\ref{line:mul-conv-carry}.
\item After all results have been computed, they are aggregated together by the call to {\tt combine} at line~\ref{line:mul-conv-combine} which computes half of the {\tt lhcs} array---the other half is computed by the second call to {\tt convolution}.
\end{itemize}

The reported code allows for all but one loops to be unrolled---the exception is the loop at line~\ref{line:mul-conv-qstart}---thus enabling scalarization of the arrays {\tt lhcs}, 
{\tt carries}, {\tt accum}.   We have found that the best performance is obtained when maximizing the size of $uint$, i.e., when $uint$ and $ubig$ are instantiated to {\tt uint64\_t} and {\tt unsigned \_\_int128}, respectively. This is as expected, since their implementation is sequentially handcrafted for that specific size, and hence offers better performance than our generic algorithm.
Finally, our implementation:
\begin{itemize}
\item utilizes two shared-memory buffers of size $IPB \cdot M \cdot \codeword{sizeof(uint)}$ bytes,
    which are necessary in order to make the elements of arrays $A$ and $B$
    available to all threads during the convolution step.
\item promotes fusion by holding the logical arrays to registers and using
    shared-memory buffers transiently for each operation. It follows that
    fusion can only be hampered by excessive register use.
\end{itemize}

\section{FFT-Based Integer Multiplication}
\label{sec:bfftmult}

This section presents our implementation for the log-linear time
integer multiplication. Several related approaches use the
domain of reals to perform the DFFT transformation, which
has the potential to affect the accuracy of the computation.  
Ensuring that errors do not manifest is often handled by restricting
the base in which the computation is carried~\cite{cuFFT-base10,JHPCA-FFT-10bits} to values that
are unfriendly to the hardware (e.g., base $10$ or $11$). 
Section~\ref{subsec:prime-fields} presents the algebraic construction
of suitable prime fields that promote efficient DFFT implementation
directly in the integral domain by exposing bases that allow near-optimal 
utilization of the underlying machine arithmetic. While this construction
is not necessarily a novelty in the computer-algebra domain, we believe
that its application to integer multiplication has merits and it is in the
least instructive to the non expert.

Finally, section~\ref{subsec:FFT-accel} sketches a relatively
straightforward implementation of the log-linear time integer
multiplication based on the Cooley-Tukey Algorithm~\cite{cooley-tukey}, which,
as before, performs an instance of multiplication with an entire 
(CUDA) block of threads, such as to allow reuse from fast memory.

\subsection{Construction of Integer Prime Fields for DFFT}
\label{subsec:prime-fields}

Finding a good prime field that enables the computation 
comes down to finding good prime numbers of shape:
\[
    p = k\cdot 2^{n} + 1
\]
that accept $2^n$ distinct roots of unity for a large enough $n$.
Applying the Little Theorem of Fermat, it follows that:
\begin{equation}
\label{little-fermat-th}
\forall a, \ \ a^{k\cdot 2^n} \ \equiv \ 1 \ \pmod p
\end{equation}
By denoting $g = a^k$ for some $a$, it follows from equation~\ref{little-fermat-th} that $g^{2^n} \equiv 1 \pmod p$. To compute a $g$ that is a $2^n$-th root of unity, one can iterate through the elements $a$ of $\mathbf{Z}_p$ and chose the first one (if any) that also verifies $g^q \neq 1, \ \forall q < 2^n$.

Once such a $g$ was successfully found, one can easily construct a $M$-th root of unity, named $\omega$, for any $M$ that is a power of $2$ and $M < 2^n$:
\[
\omega \ = \ g^{2^n/M} \ = \ g^{2^{n - \log_2(M)}}
\]
for example, one can easily check that $\omega ^ M \ \equiv \ g ^ {2^n} \ (mod~p) \ \equiv \ 1 \ (mod~p)$.

\begin{figure}[t]
\begin{lstlisting}[language=C++,style=lstFigStyle]
template<typename P> class zmod_t {
    using rep_t  = typename P::uint; 
    using ubig_t = typename P::ubig;
    static const rep_t modulus = P::p;
public:
    __host__ __device__ static rep_t norm(const rep_t v)  {
        return (0 <= v && v < modulus) ? v : v % modulus; 
    }
    __host__ __device__ static rep_t add (const rep_t x, const rep_t y) {
        ubig_t r = ((ubig_t) x) + ((ubig_t) y);
        if (r >= modulus) r -= modulus;
        return (rep_t)r;
    }
    __host__ __device__ static rep_t sub (const rep_t x, const rep_t y) {
        rep_t r = x;
        if (x < y) r += modulus;
        return (r - y);
    }
    __host__ __device__ static rep_t mul(const rep_t x, const rep_t y) {
        ubig_t r = ((ubig_t) x) * ((ubig_t) y);
        return (r % (ubig_t)modulus);
    } $\ldots$
};
\end{lstlisting}\vspace{-2ex}
\caption{ A sketch of the prime-field implementation that omits the negation, inversion and power operators. }
\label{fig:zmodp}
\end{figure}

A simple Maple program has revealed in less than $15$ minutes
of computation the following ``good'' primes that conform with the
desired shape:
\begin{description}
\item[$PrimeField32$:] $(p=3221225473, k=3, n=30, g=13)$ 
\item[$PrimeField64$:] $(p=4179340454199820289, k=29, n=57, g=21)$ 
\end{description}

$PrimeField32$ allows (i) to represent an integer (input) as an array of type $[M]uhlf$ with {\tt $uhlf$ = unsigned short}, in which only the first $15$ out of $16$ bits are utilized, (ii) the FFT computation to be carried on in extended representation $[M]uint$, in which {\tt $uint$ = uint32\_t}, and (iii) some of the prime-field computation---such as addition, multiplication, division---need to be performed in a double-sized type {$ubig$ = uint64\_t}.
Similarly, $PrimeField64$ uses the following instantiations: {\tt$uhlf$ = uint32\_t} such that only the first $31$ out of $32$ bits are utilized, {\tt$uint$ = uint64\_t} and {\tt$ubig$ = unsigned \_\_int128}. 
The implementation of several of the prime-field operations is shown in figure~\ref{fig:zmodp}.



\subsection{Straightforward Acceleration of Cooley-Tukey Algorithm}
\label{subsec:FFT-accel}

\begin{figure}
\begin{lstlisting}[language=C++,style=lstFigStyle]
template<typename P, uint32_t M, uint32_t Q> __device__ void 
fft ( typename P::uint_t* shmem, uint32_t lgM, typename P::uint_t* omegas,
      typename P::uint_t Areg[2*Q], typename P::uint_t Rreg[2*Q]
) {
    using uint_t = typename P::uint_t; using PF = zmod_t<P>;
    cpReg2Shm<uint_t,2*Q>( Arg, shmem );
    __syncthreads();
    
    for( int32_t q = 0; q < 2*Q; q++ ) {
        int32_t vtid = threadIdx.x + q*blockDim.x;
        permute<uint_t>( vtid, lgM, shmem );
    } 
    __syncthreads();
    
    for(int32_t t = 1; t <= lgM; t++) {
        uint32_t L = 1 << t,   Ld2 = L >> 1,   r = M >> t;
        for(int32_t q = 0; q < Q; q++) {
            int32_t vtid = threadIdx.x + q*blockDim.x;
            int32_t k = vtid >> (t-1);
            int32_t j = vtid &  (Ld2 - 1);
            int32_t kLj = k*L + j;
            uint_t omega_pow = omegas[r*j];
            uint_t tau       = PF::mul( omega_pow , shmem[kLj + Ld2] );
            uint_t x_kLj     = shmem[kLj];
            shmem[kLj]       = PF::add(x_kLj, tau);
            shmem[kLj + Ld2] = PF::sub(x_kLj, tau);
        }
        __syncthreads();
    }   
    cpShm2Reg<uint_t,2*Q>( shmem, Rrg );
    __syncthreads();
}

template<typename P, uint32_t M, uint32_t Q> __device__ void 
ifft( typename P::uint_t  invM,     uint32_t lgM,
      typename P::uint_t* shmem,    typename P::uint_t* omegas_inv,
      typename P::uint_t Areg[2*Q], typename P::uint_t Rreg[2*Q]
) {
    fft<P,M,Q>( shmem, lgM, omegas_inv, Areg, Rreg );
    for(int i=0; i<2*Q; i++) Rreg[i] = zmod_t<P>::mul(invM, Rreg[i]);
}

template<typename P, uint32_t M, uint32_t Q> __device__ void 
bmulFFT ( typename P::uint  invM,   uint32_t lgM,                 
          typename P::uint* omegas, typename P::uint_t* omegas_inv,
          typename P::uint* shmem,  typename P::uhlf Ahlf[Q],
          typename P::uhlf Bhlf[Q], typename P::uhlf Rhlf[Q]
) {
    using uint = typename P::uint; using uhlf = typename P::uhlf;
    uint_t Areg[Q], Afft[Q], Breg[Q], Bfft[Q], Treg[Q], Rreg[Q];

    for(int q=0; q<Q; q++) Areg[q] = Ahlf[q]; @\label{line:fft-fwd-start}@
    fft<P,M,Q/2>(shmem, lgM, omegas, Areg, Afft);
    
    for(int q=0; q<Q; q++) Breg[q] = Bhlf[q];
    fft<P,M,Q/2>(shmem, lgM, omegas, Breg, Bfft); @\label{line:fft-fwd-end}@

    for(int q=0; q<Q; q++) Treg[q] = zmod_t<P>::mul(Afft[q], Bfft[q]); @\label{line:fft-do-mult}@
    ifft<P,M,Q/2>(shmem, invM, lgM, omegas_inv, Treg, Rreg); @\label{line:fft-bwd}@

    uhlf_t Rlw[Q], Rhc[Q];
    splitFftReg<P,Q>(Rreg, (uhlf*)shmem, Rlw, Rhc); @\label{line:fft-split}@
    baddRegMul2Fft<P, M, 2*Q, 0>( (uhlf*)shmem, Rlw, Rhc, Rhlf ); @\label{line:fft-finalize}@
}
\end{lstlisting}\vspace{-3ex}
\caption{Main CUDA wrapper function that computes FFT multiplication,
        where the input ({\tt Ahlf}, {\tt Bhlf}) and result ({\tt Rhlf})
        are stored in register memory.
    }
\label{fig:fft-mul-regs}
\end{figure}

Figure~\ref{fig:fft-mul-regs} gives a relatively-straightforward CUDA implementation of the Cooley-Tukey algorithm, adapted from~\cite{vanloanfft} to use roots of unity in a finite field rather than on the unit circle in the complex plane. The entry function, denoted {\tt bmulFFT}, assumes that $M$ is a power of two and that the total sequentialization factor $Q$ is greater or equal to $2$ and it evenly divides $M$---this is ensured by suitable padding. It follows that the CUDA block size is $\frac{M}{Q}$. The arguments are: {\tt invM} is the (pre-computed) inverse of $M$ in prime field $P$, {\tt lgM} is the base-2 logarithm of $M$,  {\tt omegas} and {\tt omegas\_inv} are (pre-computed) arrays holding the $M$-roots of unity, i.e., 
\begin{lstlisting}[language=futhark,style=lstDisplayStyle]
omegas     = $\mathrm{scan}^{\mathword{exc}}$ (P::mul) 1 (replicate M $\omega$)
omegas_inv = $\mathrm{scan}^{\mathword{exc}}$ (P::mul) 1 (replicate M $\omega^{-1}$)
\end{lstlisting}
Finally, {\tt shmem} is a shared-memory staging buffer of type $[M]uint$, and
{\tt Ahlf} and {\tt Bhlf} represent the input arrays of element type $uhlf$, which were already copied in coalesced way from global to register memory. {\tt Bhlf} is the place-holder for the result and it is mapped to register memory as well.

The implementation applies FFT to the input arrays (lines~\ref{line:fft-fwd-start}-\ref{line:fft-fwd-end}), then multiplies together the FFT results within the given prime field (line~\ref{line:fft-do-mult}), and applies the inverse FFT transform (line~\ref{line:fft-bwd}), whose result is stored in {\tt Rreg}.

Function {\tt splitFftReg}, whose implementation is not shown, is called at line~\ref{line:fft-split} to change the base in which the computation is carried from $2^{8\cdot \codeword{sizeof(uint)}}$ back to $2^{8\cdot \codeword{sizeof(uhlf)} - 1}$:
\begin{itemize}
\item it first aggregates the $Q$ per-thread results of element type $uint$ into $Q$ low parts, one high part and a carry, all of element type $uhlf$; the procedure is similar to the one described in section~\ref{subsec:birds-eye-view} for classical multiplication,
\item then it places the result in a shared-memory buffer in a manner similar to the one depicted in figure~\ref{fig:place-lhcs}---except that there is no ``symmetrical'' part,
\item finally, it loads the results back again to register memory, denoted {\tt Rlw} and {\tt Rhc}---such that each thread holds the elements at the same indices from the logical arrays {\tt Rlw} and {\tt Rhc}.
\end{itemize}
The numbers represented by logical arrays {\tt Rlw} and {\tt Rhc} are finally summed up (carry propagation included) by the call to function {\tt baddRegMul2Fft} at line~\ref{line:fft-finalize}. Since the addition needs to be performed in base $2^{8\cdot \codeword{sizeof(uhlf)} - 1}$, we multiply all elements of {\tt Rlw} and {\tt Rhc} by two, perform the addition in the machine base $2^{8 \cdot \codeword{sizeof(uhlf)}}$, then divide back by $2$ while reapplying the carries. (An element resulted from addition was subject to a carry if its value is odd.)

The implementation of {\tt fft} and {\tt ifft} are standard. 
FFT-based multiplication can operate with integers having (close to) the same size as the classical multiplication ($\mathcal{CM}$): on the one hand its input uses half the element type of $\mathcal{CM}$ (i.e., $uhlf$ {\em vs.} $uint$), but on the other hand it uses only one memory buffer of size $M\cdot \codeword{sizeof(uint)}$ instead of two buffers with $\mathcal{CM}$.  To be exact, the maximal supported size is a bit smaller than the one of $\mathcal{CM}$, because only $15$ out of $16$ bits can be used with $PrimeField32$ and $31$ out of $32$ bits with $PrimeField64$.


\section{Futhark's Strengths and Weaknesses}
\label{sec:futhark}

\subsection{Futhark Optimizations Relevant for Big Integer Arithmetic}
\label{subsec:futhark-opts}
\subsubsection{Incremental Flattening.}
Futhark supports expression of parallel programs that operate on regular multi-dimensional arrays. The arbitrarily-nested application parallelism is flattened~\cite{blelloch1990vector,blelloch1994implementation}, by a technique dubbed ``incremental flattening''~\cite{futhark-ppopp} that utilizes map fission and map-loop interchange to create semantically-equivalent code versions that systematically map more and more levels of application parallelism to the hardware.
Essentially, when a new \lstinline{map f} operation is discovered in the top-down traversal of the program, the analysis:
\begin{itemize}
\item[(1)] creates a first code version corresponding to a CUDA kernel in which each thread executes (independently) an application of $f$.
\item[(2)] creates a second code version in which the \lstinline{map} parallelism discovered so far is mapped on the CUDA grid, and the parallelism inside function $f$ is recursively flattened and mapped to CUDA block level, such that intermediate arrays are mapped to shared memory. This is dubbed an {\em intra-group kernel}.
\item[(3)] the current \lstinline{map} is added to the parallel context---that represents a perfect nest of \lstinline{map} operations---and incremental flattening continues recursively.
\end{itemize}

The resulted code versions are independently optimized and combined into one program by guarding each of them with a predicate that compares compare a dynamic program measure\footnote{In practice, the dynamic measure is the degree of parallelism utilized by a kernel.} with a threshold. Threshold values are autotuned---so as to select the best combination of code versions---based on deterministic procedure that is guaranteed to produce a near-optimal result in minimal number of runs, as long as the dynamic measure conforms with a monotonic property~\cite{futhark-tuning}.

This compilation strategy has proved valuable in enabling high-performance implementation of real-world applications from domains such as  computer vision~\cite{annf}, remote-sensing~\cite{bfast,STL-rem-sens} and computational finance~\cite{LexiFiPricing}, and, as well, in optimizing code synthesized by program-level transformations such as automatic differentiation~\cite{futhark-ad-sc22,ifl-rev-ad}.
Similarly, we have found that incremental flattening is essential to supporting multi-precision computations in an high-level architecture-neutral language such as Futhark: On the one hand, choice (2) produces the intra-group kernel that we are aiming for, i.e., that leverages the use of fast (shared) memory and supports efficient fusion of such operations. One the other hand, it provides a fail-safe platform: the computation will still be carried on, albeit less efficiently so, on large integers that do not fit the intra-group kernel, by means of the versions generated by choices (3) and (1). 

\subsubsection{Memory Optimizations.} A set of analyses that come handy in our context refer to reducing the memory footprint~\cite{philip-thesis} and at eliminating unnecessary copy operations, dubbed short-circuiting analysis~\cite{futhark-mem-sc22}.  The former corresponds to applying register-like allocation to operate on memory buffers instead of registers, thus allowing buffers' reuse once their liveness ended. This reduces the shared-memory requirements of the kernel and enables larger integers and fusion.

Short-circuiting analysis~\cite{futhark-mem-sc22} addresses an inefficiency common to functional languages whose type systems enforce correct-by-construction parallelism: some parallel loops---e.g., appearing in LUD or  Needleman-Wunsch algorithms---cannot be expressed directly because they both read and write (non-overlapping slices of) elements of the same matrix. The typical type system does not perform dependence-analysis on arrays~\cite{SummaryMonot,CIVan} and will demand to separate the parallel loop into two parallel operations, typically:  a \lstinline{map} that reads from the original matrix and results into a temporary buffer, and another parallel-write operation that updates the corresponding slice of the matrix with the buffer elements.  

The analysis introduces a notion of memory and attempts to map the memory space of the buffer directly to the corresponding memory space of the matrix---whenever it can guarantee safety---such that the parallel write becomes a {\tt noop} and the buffer does not actually allocates any extra space. A trivial example is:
\[
\kw{let} \ x \ = \ \kw{concat} \ a \ b
\]
If $a$ or $b$ are lastly used in the above statement, then their memory is allocated directly in the corresponding memory space of $x$ and the \lstinline{concat} becomes a {\tt noop}.

\begin{figure}[t]
\begin{lstlisting}[language=futhark,style=lstFigStyle]
let badd [ipb][n] (as: [ipb*(4*n)]u32) (bs: [ipb*(4*n)]u32) : [ipb*(4*n)]u32 =
  let g = ipb * n
  let cpGlb2Sh (i : i64) = #[unsafe]
        ( ( as[i], as[g + i], as[2*g + i], as[3*g + i] )
        , ( bs[i], bs[g + i], bs[2*g + i], bs[3*g + i] ) )

  let ( ass, bss ) = map cpGlb2Sh (0...<g) |> unzip @\label{line:fut-eff-read}@
  let (a1s, a2s, a3s, a4s) = unzip4 ass
  let (b1s, b2s, b3s, b4s) = unzip4 bss
  let ash = a1s ++ a2s ++ a3s ++ a4s @\label{line:fut-eff-concat1}@
  let bsh = b1s ++ b2s ++ b3s ++ b4s @\label{line:fut-eff-concat2}@
  ...
\end{lstlisting}\vspace{-2ex}
\caption{Futhark code illustrating coalesced copying from global to shared memory for an efficient sequentialization factor $Q=4$.
}
\label{fig:futhark-eff-seq}
\end{figure}

We have used this compiler feature to support efficient sequentialization in our Futhark implementations. Figure~\ref{fig:futhark-eff-seq} shows the prelude of the function that performs integer addition, which is intended to be mapped at CUDA block level.  The \lstinline{map cpGlb2Sh} operation on line~\ref{line:fut-eff-read} reads four elements of {\tt as} and {\tt bs} with each of the {\tt g} threads (of the block) in coalesced way from global to shared memory---i.e., consecutive threads access consecutive words in global memory. The following \lstinline{concat} operations at lines~\ref{line:fut-eff-concat1} and~\ref{line:fut-eff-concat2} put the results together in the correct order in arrays {\tt ash} and {\tt bsh}, which will be mapped to shared memory. Short-circuiting analysis ensures that only two shared-memory buffers are allocated (for the final {\tt ash} and {\tt bsh}), and that the \lstinline{concat} operation cost nothing---since {\tt{}a1s$\ldots$a4s} are allocated directly in the memory space of {\tt ash}. 

\subsection{Shortcomings of Futhark's Compiler Infrastructure}
\label{subsec:futhark-lacks}

The experimental evaluation, reported in the next section, shows that 
our Futhark implementation has sub-optimal performance and scalability in comparison to our \cuda~prototype and the \cgbn~library. The central reason is the absence of a compiler pass aimed at supporting efficient sequentialization. Rationale is:

{\em First}, performing efficient sequentialization by hand is not only ``un-natural'' and results in less elegant code, but, more importantly it has the potential of degrading the performance of the other semantically-equivalent code versions. 
For example, if the integer size is too large to fit in the intra-group kernel, then the code in figure~\ref{fig:futhark-eff-seq}---which was intended to copy in coalesced way from global to shared memory---performs in a convoluted way two expensive and completely unnecessary copies (global-to-global memory).

{\em Second}, logical array created inside the intra-group kernel are currently mapped by the compiler to shared-memory only, since this guarantees that their elements are accessible to any threads. This mapping has serios performance implications since shared memory has higher latency than registers, and there is no manner in which this mapping can be altered by the programmer. It also follows that Futhark kernels will requires more shared memory than necessary, which limits the magnitude of the supported integer. For example, in the case of addition, our \cuda~ prototype utilizes one memory buffer (and this can be further reduced), while Futhark requires twice as much.

{\em Third}, the Futhark FFT implementation currently uses a \lstinline{scatter} (parallel write) operation as the result of a loop (body), which requires two shared-memory buffers that are aliased across the loop (double buffering).  This circular aliasing prevents the current compiler to reuse the space of these buffers for subsequent operations, even when their liveness interval has ended. We observe that if the corresponding array was mapped to register memory instead, then only one shared-memory buffer would be necessary, and furthermore the double buffering of that array would be efficiently supported by the register allocation of the underlying compiler (e.g., {\tt nvcc}).

{\em Finally}, Futhark does not yet supports an $128$-bit integer, which would offer a significant boost to the performance of the classical multiplication.

\section{Experiments}
\label{sec:experiments}

This section evaluates the performance of our implementations for addition, multiplication and fusion of such operations. We compare our results with those of the ``Cooperative Groups Big Numbers'' library,\footnote{
\href{https://github.com/NVlabs/CGBN}{https://github.com/NVlabs/CGBN}
} (\cgbn) authored by NVlabs. that offers a framework for performing unsigned multiple precision integer arithmetic in CUDA. The current release (XMP 2.0 Beta) offers state-of-the-art performance on small to medium sized integers: $2^5$ bits through $2^{15}$ bits, but also seems to support larger integers, albeit without top-performance guarantees. 

The key design decision in \cgbn~is that one operation is performed within (at most) one warp of threads, such that the implementation can leverage specialized hardware (instructions) that enable very efficient (low-latency) communication of register values within a warp. This also promotes the scalability of fused operations.
In comparison, our implementations does not rely on specialized hardware instructions, and maps an integer instance to be solved by at most one CUDA block of threads.  We thus expect to achieve higher performance on larger integers, where \cgbn~is likely to be affected by high-register pressure.

\subsection{Hardware, Benchmark, Performance Measures, Methodology}
\label{subsec:method}

Our evaluation uses an Nvidia A100 GPU that offers $6912$ cores, 
peak global-memory bandwidth of $1.555$ TB/sec and {\tt FP32} peak
performance of $19.49$ Tflops.

We evaluate our implementations in comparison with \cgbn~by running programs
that perform batches of (i) one addition \textbf{1-Add}, (ii) six additions \textbf{6-Add}, (iii) one multiplication \textbf{1-Mul}, and (iv) a polynomial computation \textbf{Poly} involving four multiplications and two additions. All four programs corresponds to the execution of one kernel such that related sequences of additions and multiplications are performed inside the same CUDA block---i.e., \textbf{6-ADD} and \textbf{Poly} are intended to evaluate the scalability of block-level fusion, in which intermediate results are maintained in fast memory.
These programs are evaluated on eight combinations of values for the size in bits $\mathword{NumBits}$ of the integer and the total number of (integer) instances $\mathword{NumInsts}$, such that $\mathword{NumBits} \cdot \mathword{NumInsts}  =  2^{32}$. 

We report the performance of
\begin{description}
\item[addition:] in GB/sec---because addition it is memory bound---and we compute
the number of bytes accessed for both \textbf{1-Add} and \textbf{6-Add} with
the formula:\footnote{
Ideally, both programs read two integers from global memory and write one as result.
}
\[
\mbox{\tt number-of-bytes-accessed} \ = \ 3 \mathword{NumInsts} \cdot \frac{\mathword{NumBits}}{8}
\]
\item[multiplication:] in terms of Giga 32-bit unit operations per second 
        (Gu32ops/sec), since multiplication is compute bound.  
        \textbf{1-Mul} uses as number of operations:
        \[
        \mbox{\tt 1-Mul-num-u32-ops} = 300 \cdot \mathword{NumInsts} \cdot m \cdot \log m, \  \mbox{where} \ m = \frac{M\cdot \codeword{sizeof(uint)}}{4}
        \]
        For \textbf{Poly} we consider the number of unit operations to be
        four times that of \textbf{1-Mul}, i.e., we only consider the four 
        multiplications and ignore the two additions.
        The rationale for the constant $300$ is that the algorithm performs three
        FFT transformations, each of them using about $100 \cdot m \cdot \log m$
        unit operations.\footnote{ We have used test programs to measure the latency 
        of 32 and 64 bit operations such as addition, multiplication, modulo, in comparison with 32-bit integer addition (as the unit), and we have counted 
        that FFT multiplication instantiated to {\tt FftPrime32} requires about 
        $100$ units inside its $M \cdot \log M$ loop nest. 
        In fact the modulo operation on {\tt uint64\_t} alone accounts for about
        $78$ additions.
        } 
        However, the constant does not matter much: the key is that the measure 
        implements a normalized runtime that allows meaningful comparison across
        different implementations and also across different datasets. 
\end{description}

Since all Cuda programs consists of one kernel call, we measure the runtime as the average of $500$ kernel runs for \textbf{1-Add}, \textbf{6-Add} and \textbf{1-Mul} and of $125$ runs for \textbf{Poly}. For Futhark, we use the option {\tt bench --backend=cuda} that (i) measures all overheads except for device initialization, kernel compilation and data transfers between host and device, and (ii) performs enough runs until the $95\%$-confidence percentile average stabilizes.

\subsection{Performance of Addition}
\label{subsec:add-perf}

Table~\ref{tab-add} shows the performance of integer addition expressed as memory bandwidth, i.e., in GB/sec. The peak global-memory bandwidth of the Nvidia A100 hardware is $1.555$TB/sec. The first two columns correspond to the number of bits of the integer ($\mathword{NumBits}$) and the total number of instances performed $\mathword{NumInsts}$. The columns denoted  \cgbn~ correspond to the performance of the \cgbn ~library, while the columns denoted by \textbf{Our-Cuda} and \textbf{Futhark} correspond to the performance of our Cuda and Futhark implementations. 

For \cgbn~ we set the thread-per-instance parameter for $\mathword{NumBits}$ equal to $2^{11}$ and $2^{12}$ to $16$ and $8$, respectively, and to $32$ for the rest of $\mathword{NumBits}$ values; we have observed that best performance is achieved for these instantiations (for both addition and multiplication).  For our Cuda and Futhark implementations, we instantiate $uint$ to {\tt uint64\_t} and the sequentialization factor $Q$ to $4$, i.e., each thread computes $4\cdot 64 = 256$ bits sequentially.

\begin{table}[t]
    \caption{Performance of Addition in \textbf{GB/sec}.
                A100's peak bandwidth is \textbf{1555} GB/s.}
    \label{tab-add}
\begin{center}
\begin{tabular}{|c|c||r|c|c||r|c|c|}
    \hline
    \textbf{Num} & \textbf{Num} & \textbf{1-Add} & \textbf{1-Add} & \textbf{1-Add} & \textbf{6-Add} & \textbf{6-Add} & \textbf{6-Add}\\
    \textbf{Bits}& \textbf{Insts}& \cgbn  & \textbf{Our-Cuda} & \textbf{Futhark} & \cgbn  & \textbf{Our-Cuda} & \textbf{Futhark}\\
    \hline\hline
    $2^{18}$ & $2^{14}$ & 369 & \textbf{1320} & 737   & 362   &  \textbf{570} &  294 \\\hline 
    $2^{17}$ & $2^{15}$ & 368  &  \textbf{1331}  & 1172  & 353   &  \textbf{803} &  350 \\\hline 
    $2^{16}$ & $2^{16}$ & 376  &  \textbf{1358}  & \textbf{1343}  & 353   &  \textbf{853} &  313 \\\hline
    $2^{15}$ & $2^{17}$ & 329  &  \textbf{1363} & \textbf{1363}  & 321   &  \textbf{856} &  431 \\\hline
    $2^{14}$ & $2^{18}$ & 581  &  \textbf{1334}  & \textbf{1370}  & 546   &  \textbf{836} &  434 \\\hline
    $2^{13}$ & $2^{19}$ & 1238 &  \textbf{1350}  & \textbf{1364}  & \textbf{1207}  &  816 &  435 \\\hline
    $2^{12}$ & $2^{20}$ & \textbf{1329} &  \textbf{1359}  & \textbf{1364}  & \textbf{1189}  &  856 &  435 \\\hline
    $2^{11}$ & $2^{21}$ & 1275 &  \textbf{1366} & \textbf{1364} & \textbf{1167}  &  855 &  435 \\\hline
\end{tabular}
\end{center}
\end{table}

Key observations derived from table~\ref{tab-add} are:
\begin{description}
    \item[\textbf{1-Add}:] both our Cuda and Futhark implementations outperform CGBN  on integers whose number of bits are in the interval $2^{14} \ldots 2^{18}$. For example, CGBN commonly achieves less than $25\%$ of the peak bandwidth, while our implementations commonly achieve higher than $85\%$ peak performance. CGBN offers competitive performance on $\mathword{NumBits}=2^{11} \ldots 2^{13}$.
    \item[\textbf{6-Add}:] CGBN offers near-perfect scalability, i.e., it takes about the same amount of time to perform six addition as it takes to perform one. We attribute this to the low-latency of specialized-register instructions for transferring values within a warp of threads. However, while \textbf{\textsc{cgbn}} offers excellent performance on integers of size $2^{11} \ldots 2^{13}$ bits (up to $1.5\times$ faster than ours), \textbf{Our-Cuda} still holds the upper hand on sizes $2^{14} \ldots 2^{18}$ (up to $2.7\times$ faster than CGBN).
    \item[\textbf{6-Add}:] our CUDA implementation offers decent scaling: except for $2^{18}$ bits, computing six additions takes less than $1.65\times$ the time of one addition. 
    \item[\textbf{6-Add}:] the scalability of the Futhark implementation is severely handicapped by the layout that maps intermediate arrays in shared-memory buffers rather than registers: six additions require (more than) $3\times$ the time of one addition. 
\end{description}

\subsection{Performance of Multiplication}
\label{subsec:mul-perf}

Table~\ref{tab-mul} shows the performance of multiplication expressed in Gu32ops/sec (see section~\ref{subsec:method}), where the best two numbers are displayed in bold text---the higher the number the better the performance.  Cells filled with \illegal~denote that kernel launch failed due to out of resources.\footnote{We prevented Futhark from switching to the slower versions described in section~\ref{subsec:futhark-opts}.} As before, the first two columns report the integer size in bits and the number of instances performed, and the columns denoted  \cgbn~ correspond to the performance of the CGBN library.  The columns denoted \cudaQuad~ and \cudaFft~ correspond to our CUDA implementation for the classical (quadratic time) and FFT (log-linear time), respectively, and similarly, the columns denoted \futharkQuad~ and \futharkFft{}~correspond to Futhark. 

\begin{table}[t]
    \caption{Performance of multiplication in Gu32ops/sec; the number of 32-bit operations 
                for \textbf{1-Mul} is computed as $300\cdot \mathword{NumInsts}\cdot m \cdot \log m$, where $m = \frac{M\cdot \codeword{sizeof(uint)}}{4}$. \textbf{Poly} computes $(a \cdot a + b)  \cdot  (b \cdot b + b) + a \cdot b$ using four multiplications and two additions; its number of operations is considered to be four times that of \textbf{1-Mul}.
            }
    \label{tab-mul}
\begin{center}
\begin{tabular}{|c|c||r|r|r|c|c||r|r|r|c|c|}
    \hline
    Num & Num & \textbf{1-Mul} & \textbf{1-Mul} & \textbf{1-Mul} & \textbf{1-Mul} & \textbf{1-Mul} & \textbf{Poly} & \textbf{Poly} & \textbf{Poly} & \textbf{Poly} & \textbf{Poly} \\
    \textbf{Bits}& \textbf{Insts} & \cgbn & \cudaQuad & \futharkQuad & \cudaFft & \futharkFft & \cgbn  & \cudaQuad & \futharkQuad & \cudaFft & \futharkFft \\
    \hline\hline
    $2^{18}$ & $2^{14}$ &   72  &\illegal & \textbf{1813}  & \textbf{11590}&\illegal &  49   &\illegal & \textbf{1795} & \textbf{11351}&\illegal \\\hline 
    $2^{17}$ & $2^{15}$ &  997  &  \textbf{4471}   & 3296  & \textbf{11789}&\illegal & 192   &  \textbf{4027}   & 3259  & \textbf{12130}&\illegal \\\hline 
    $2^{16}$ & $2^{16}$ &  6482 &  \textbf{7843} &  5901 & \textbf{11466}&\illegal & 5837  &  \textbf{7148}   & 5820  & \textbf{11679}&\illegal \\\hline
    $2^{15}$ & $2^{17}$ &  11640&  \textbf{13460} & 10091 & \textbf{12779}& 10187& \textbf{12246} &  \textbf{12247} & 9889  & \textbf{12621}& 8395 \\\hline
    $2^{14}$ & $2^{18}$ & \textbf{21461}&  \textbf{21608}& 15856 & 14297& 11259& \textbf{21454} &  \textbf{19455}  & 15093 & 13899& 10742\\\hline
    $2^{13}$ & $2^{19}$ &  \textbf{34004}&  \textbf{31658}  & 24267 & 15791& 11651& \textbf{34165} & \textbf{27876}  & 20677 & 15173& 11620\\\hline
    $2^{12}$ & $2^{20}$ &  \textbf{54465} &  \textbf{49328}  & 39861 & 15264& 11051& \textbf{53846} &  \textbf{42673}  & 30569 & 11189& 9641 \\\hline
    $2^{11}$ & $2^{21}$ &  \textbf{86252}&  \textbf{70661}  & 61333 & 13554& 10099& \textbf{86182} &  \textbf{60819}  & 44094 & 10884& 8801 \\\hline
\end{tabular}
\end{center}
\end{table}

Our implementation of classical multiplication specializes $uint$ to {\tt uint64\_t} and use a total sequentialization factor of $4$, i.e., each thread computes two elements from the first half and their two symmetric opposites across the middle, as in figure~\ref{fig:place-lhcs}.
For FFT multiplication we use (i) the smallest sequentialization factor (greater than two) that allows the computation to fit in a CUDA block, which is constrained to $1024$ threads, and (ii) the {\tt FftPrime32} field, in which $uint$ is {\tt uint32\_t} and $ubig$ is {\tt uint64\_t}. Using {\tt FftPrime64} is a bit slower mainly because it requires a modulo operation on $128$-bit integers, which is very expensive.
%
%
Key observations derived from table~\ref{tab-mul} are:
\begin{itemize}
    \item[(1)] On both \textbf{1-Mul} and \textbf{Poly}, our CUDA implementation of classical (quadratic) multiplication is faster than \cgbn~ for integer sizes in the range $2^{15} \ldots 2^{18}$. Size $2^{15}$ is also very close to the split point from which on, our CUDA FFT implementation starts outperforming the quadratic implementation. 
    \item[(2)] \cgbn~is faster on the smaller integer sizes $2^{11} \ldots 2^{13}$ by factors as high as $1.2\times$ and $1.4\times$ on \textbf{1-Mul} and \textbf{Poly}, respectively, but our \cudaQuad~is faster on sizes $2^{15} \ldots 2^{17}$ by factors as high as $4.5\times$ and $21\times$.
    \item[(3)] Our CUDA FFT implementation is faster than the best quadratic running implementation (i) by factors of $6.3\times$ on integer size $2^{18}$, and (ii) by $2.6\times$ and $3.0\times$ factors on integer size $2^{17}$ on \textbf{1-Mul} and \textbf{Poly}, respectively.
    \item[(4)] \cgbn~demonstrates excellent (super-linear) scalability on integer sizes between $2^{11} \ldots 2^{15}$, since its performance on \textbf{Poly} is very-close to or better than the one on \textbf{1-Mul}, even when the two additions are not counted.
    \item[(5)] Our CUDA FFT implementation also demonstrates excellent scalability on sizes higher than or equal to $2^{15}$, which is all that matters because $2^{15}$ seems to be the split point from which point on FFT gains the upper hand.
    \item[(6)] We attribute the performance gap between our CUDA and Futhark quadratic implementations on \textbf{1-Mul} to the fact that Futhark lacks support for $128$-bit integers, and hence it uses less-efficient $64$-bit arithmetic that computes the high and low parts. Similar to addition, Futhark's scalability (for \textbf{Poly}) is worse due to logical arrays being always mapped to shared memory.
    \item[(7)] Finally, the Futhark FFT implementation runs out of resource for sizes of $2^{16}$ to $2^{18}$ bits, due to the last issue reported in section~\ref{subsec:futhark-lacks}.
\end{itemize}

\section{Conclusions}
\label{sec:conclusions}

We have shown that level languages (\texttt{C++} and Futhark) can be used to implement big integer addition and multiplication concisely and efficiently for GPU computation.  These implementations are simple and efficient for big integers of practical size,
comparing favourably to the \cgbn~library for integers of size from $2^{15}$ to $2^{18}$ bits (i.e. up to about 79,000 digits).
We have seen that an FFT-based multiplication can, by factors as high as $5\times$, outperform  an efficient implementation of the classical multiplication on sizes that fit in a \cuda~block.   This is achieved using a naive implementation of finite field arithmetic --- further improvement would be expected using Montgomery representation.
The paper has presented the implementation in sufficient detail to be reproduced as desired by others.   For \texttt{C++} template CUDA code has been provided.   

We have measured the performance of these implementations against the high quality \cgbn{} library, testing up to sizes of $2^{18}$ bits. 
For addition, our CUDA code outperforms the \cgbn{} library by a factor of about $2\times$ to $4\times$ for integers of more than about $2^{14}$ bits.   For most tests, our functional Futhark code also outperforms the \cgbn{} library.   
For classical quadratic multiplication, our simple CUDA code is comparable to the \cgbn{} library for numbers with upto $2^{14}$ bits and superior to \cgbn{} for larger sizes.   The Futhark implementation is comparable over most of the range of sizes.   
The CUDA FFT implementation of multiplication is superior for sizes greater than $2^{15}$ bits, becoming about 160 times faster at $2^{18}$ bits.
For tests involving a combination of operations (the ``Poly'' tests),  our CUDA implementation using classical arithmetic performs significantly better than \cgbn{} for sizes above $2^{15}$ bits and within a factor of 2 below that size.  While the Futhark implementation meets the criterion of being concise and flexible, furhter compiler support is required to approach the efficiency of our CUDA code.  The present investigation has identified specific areas of Futhark compiler enhancement that together may lead to performance comparable to our CUDA code.

\subsubsection{\ackname} The authors were originally inspired to consider non-uniform memory architectures in a collaboration with Alan Mycroft~\cite{OanceaMycroftWatt,OanceaSetCongrDynAn}, without whom we never would have ended up here!

\bibliographystyle{splncs04}
\bibliography{references.bib}

%
%
%
%
%
%
%
%
%

\end{document}